\documentclass[12pt,preprint]{aastex}

\newtheorem{theorem}{Theorem}
\newtheorem{lemma}{Lemma}

\newenvironment{proof}{~ \\[0.1in] {\bf Proof.} }{\hfill $\Box$ \bigskip \\[0.1in]}

\usepackage[intlimits]{amsmath}
\usepackage{times}

\newcommand{\disp}{\delta}
\newcommand{\aone}{a}
\newcommand{\atwo}{\tilde{a}}
\newcommand{\xt}{\tilde{x}}
\newcommand{\yt}{\tilde{y}}
\newcommand{\rt}{\tilde{r}}
\newcommand{\Et}{\tilde{E}}
\newcommand{\Ft}{\tilde{F}}
\newcommand{\It}{\tilde{I}}
\newcommand{\htild}{\tilde{h}}
\newcommand{\Ht}{\tilde{H}}
\newcommand{\Rt}{\tilde{R}}
\newcommand{\Mone}{{\cal M}_1}
\newcommand{\Mtwo}{{\cal M}_2}
\newcommand{\Mthree}{{\cal M}_3}
\newcommand{\Mfour}{{\cal M}_4}
\newcommand{\thetat}{\tilde{\theta}}
\newcommand{\Thetat}{\tilde{\Theta}}
\newcommand{\Dxt}{\Delta\tilde{x}}
\newcommand{\xtt}{\tilde{\tilde{x}}}
\newcommand{\dxt}{\Delta\tilde{x}}
\newcommand{\phit}{\tilde{\phi}}
\newcommand{\phitt}{\tilde{\tilde{\phi}}}

\newcommand{\xb}{\hat{x}}

\shorttitle{Pupil Mapping in 2-D}
\shortauthors{Vanderbei and Traub}

\begin{document}

\title{Pupil Mapping in 2-D for High-Contrast Imaging}

\author{Robert J. Vanderbei}
\affil{Operations Research and Financial Engineering, Princeton University}
\email{rvdb@princeton.edu}

\and

\author{Wesley A. Traub}
\affil{Harvard-Smithsonian Center for Astrophysics}
\email{wtraub@cfa.harvard.edu}

\begin{abstract} 
Pupil-mapping is a technique whereby a uniformly-illuminated
input pupil, such as from starlight, can be mapped into a non-uniformly
illuminated exit pupil, such that the image formed from this pupil will have
suppressed sidelobes, many orders of magnitude weaker than classical Airy ring
intensities.  Pupil mapping is therefore a candidate technique for
coronagraphic imaging of extrasolar planets around nearby stars.  
The pupil mapping technique is lossless, and preserves the full 
angular resolution of the collecting telescope, so it could 
possibly give the highest signal-to-noise ratio of any proposed 
single-telescope system for detecting extrasolar planets.
A planet fainter than $10^{-10}$ times its parent star, and as close as 
about $2 \lambda/D$ should be detectable.  We derive
the 2-dimensional equations of pupil mapping for both 2-mirror and 2-lens
systems.  We give examples for both cases.  We derive analytical estimates of
aberration in a 2-mirror system, and show that the aberrations are essentially
corrected with an added reversed set of mirrors.  
\end{abstract}


\section{Introduction} \label{sec:intro}

To successfully detect and characterize extrasolar terrestrial planets around
nearby stars, it is necessary to isolate the light of a planet from that of
the parent star to better than about $10^{-10}$ at visible wavelengths 
or $10^{-6}$ at
thermal infrared wavelengths (\cite{ref:DHJKLLSSTW02}).  
Two general types of
space-based instruments have been proposed to do this, visible coronagraphs
and infrared interferometers.  

At present, NASA plans to launch the Terrestrial Planet Finder-Coronagraph
(TPF-C) in 2014, and the Terrestrial Planet Finder-Interferometer (TPF-I) 
in 2019.
ESA plans to launch an infrared interferometer, DARWIN, around 2015.  Each of
these observatories appears to be feasible using current or expected
technology, and for each there are several alternative architectures now under
study.  The present paper discusses the underlying mathematical principle of
the coronagraph concept known as {\em pupil mapping} (\cite{TV03}),
alternatively called {\em intrinsic apodization} (\cite{GOPP02}) or
{\em phase-induced amplitude apodization (PIAA)} (\cite{Guy03}, \cite{GGRSO04}).  

The basic idea of pupil mapping is that the uniform intensity of starlight
falling on the input pupil of a telescope can be mapped, ray by ray, to a
non-uniform intensity in an exit pupil, such that the image of a star will be
highly concentrated with minimal sidelobes.  The goal is to reduce the
sidelobes to less than $10^{-10}$ within a very few diffraction widths from the
central star.  This permits us to separate starlight from planet-light.  For
perspective, sidelobes less than $10^{-10}$ are roughly 8 orders of magnitude 
lower than the Airy-ring pattern that would be expected from an idealized
conventional telescope image.

Several other coronagraphic methods have been suggested.  One of the first was
the idea of a square apodized pupil (\cite{ref:Nisenson}), in which
the transmission function of the pupil is tapered to zero at the edges,
thereby reducing the sidelobes, but at a loss of light and angular resolution.
Another idea is the shaped pupil (\cite{ref:Spergel}, 
\cite{VKS03,VSK03,VSK02,VKS04}), 
in which the pupil is covered by an opaque mask that has
carefully-shaped transmitting cut-outs, thereby strongly reducing the
sidelobes in two azimuthally-opposite segments, but again at a loss of light
as well as some loss of angular resolution.  A different approach was taken by
\citet{ref:kuchner}, \citet{ref:KS03a}, and \cite{KCG04}, 
who proposed a family of image-plane band-limited masks that would
block starlight and transmit planet light, with little loss of planet light
and nearly full angular resolution.  Yet another idea was suggested by
\cite{LSLWL03} who combine pupil-shearing with single-mode fibers,
a combination that also potentially has good transmission and angular
resolution.  Other methods have also been proposed, some of which are more
limited in spectral range; see, for example, \cite{AS03}.

Even before Guyon first suggested pupil mapping for TPF-C, there was an
abundant literature on the topic of beam-shaping, in both the radio astronomy
community (in order to optimally couple a telescope beam into a detector
horn), and in the laser community (to reshape the Gaussian beam from a laser
into a more generally useful uniform-intensity beam).  The laser beam shaping
is closely related to the present pupil mapping for astronomy, because the
laser beam work particularly aimed at maintaining a flat wavefront after the
shaping optics.  In particular we mention the U.S. patent awarded to 
\citet{Kreuzerpatent}, and a more recent, 
but representative, application of this work by
\cite{HJ03}.  In fact, one of the equations that we derive
in the present paper (equations \eqref{430} and \eqref{431} in Theorem
\ref{thm2}), for the case of pupil mapping with lenses, is
identical to that derived by Kreuzer, the only difference being that the
latter was trying to make a uniform-intensity beam from a Gaussian one, and we
are trying to do essentially the reverse of that.  

The present paper is an extension of our previous one (\cite{TV03}), 
in that we move from an idealized 1-D treatment to the more physical 2-D
case, in addition to which we now also present equations for pupil mapping
with lenses as well as mirrors.  We furthermore include an analytical
development of off-axis aberrations, and a method for removing the aberrations
of off-axis images.  We give several illustrative examples of pupil mapping,
including a section showing how one simple type of pupil-mapping can directly
yield the entire family of Cassegrain and Gregorian paraboloid-based
telescopes, including the simple plane-mirror periscope as a special case.

Our approach in this paper is the same as in \citet{TV03},
namely to develop the analytical basis of pupil mapping, and explore the
consequences.  For some questions, including that of fabrication, it will be
necessary to carry out numerical investigations, which we defer to a future
paper.

The organization of the present paper is as follows.  
In Section \ref{sec:pm}, we define the ray-by-ray mapping function $R$.
In Section \ref{sec:ms}, we derive the differential equations describing the
mirror surfaces in terms of $R$.
In Section \ref{sec:ea}, we show how $R$ can be calculated from the function
$A^2$, where $A^2$ is the point-by-point ratio of output to input intensity
across the pupil.
In Section \ref{sec:ex}, we give three examples of pupil-mapping, for
constant, Gaussian, and prolate-spheroidal-like functions.
In Section \ref{sec:ext}, we extend the theory to include on-axis mirror
systems.
In Section \ref{sec:refract}, we extend the theory to 2-lens systems.
In Section \ref{sec:ellip}, we show how to modify the equations to apply
to elliptically shaped pupils.
In Section \ref{sec:oa}, we estimate the magnitude of off-axis aberrations in
2-mirror systems and show that the aberrations are nearly eliminated with a
pair of identical but reversed mirrors.

\section{Pupil Mapping} \label{sec:pm}

Consider a $2$-mirror optical pupil-mapping system endowed with a 
Cartesian coordinate system in which
the $z$-axis corresponds to the optical axis so that the pupil and
image planes are
parallel to the $(x,y)$-plane.  The first mirror's projection onto the
$(x,y)$-plane is a circle of radius $\aone$ centered at $(0,0)$:
\begin{equation}\label{1}
     \Mone = \{ (x,y): x^2 + y^2 \le \aone^2 \} .
\end{equation}
The second mirror has radius $\atwo$ and is shifted along the $x$-axis by
a distance $\delta$.
Its projection on the $(\xt,\yt)$-plane is
\begin{equation}\label{2}
     \Mtwo = \{ (\xt,\yt): (\xt-\delta)^2 + \yt^2 \le \atwo^2 \} .
\end{equation}
The displacement $\delta$ can be any real value, but if the mirrors are
nonoverlapping then we need
\begin{equation} \label{650}
    |\delta| \ge \aone + \atwo .
\end{equation}
If $\delta = 0$, then the optics are concentric.  We show two such examples;
for mirrors in Section \ref{sec:conc}, and for lenses in Section
\ref{sec:refract}.
We introduce polar coordinates $r$ and $\theta$ for $\Mone$,
and $\rt$ and $\theta$ for $\Mtwo$.
Hence, 
\begin{eqnarray}
    x = r \cos \theta, & \qquad & y = r \sin \theta \label{3} \\
    \xt = \rt \cos \theta + \disp, & \qquad & \yt = \rt \sin \theta \label{4} .
\end{eqnarray}
Note that $r=0$ and $\rt = 0$ refer to physically different locations (the
centers of $\Mone$ and $\Mtwo$, respectively), but $x=0$ and $\xt = 0$ are the
same physical location.
The mirror surfaces are at
$z = h(x,y)$ for the first mirror
and $\tilde{z} = \htild(\xt,\yt)$ for the second mirror.
Light enters the system from above, reflects upward off from the first mirror
and then impinges on the second mirror, which reflects it back downward.

A pupil mapping is determined by specifying a one-to-one
and onto mapping between the
two mirrors, or equivalently between their two projections $\Mone$ and
$\Mtwo$.  In general, such mappings could be rather elaborate.  To keep
the design (and analysis) simple, we assume that polar angle $\theta$ on
the first mirror maps to the same polar angle on the second one.  
Hence, the pupil
mapping is completely determined by giving a function $\Rt$ from 
mirror-one radii to mirror-two radii.  Hence, we have
\begin{equation}\label{5}
    \rt = \Rt(r) .
\end{equation}
To be one-to-one, $\Rt$ must map $[0,\aone]$ monotonically onto $[0,\atwo]$.
In such a case, there is an inverse function $R$:
\begin{equation}\label{6}
    r = R(\rt) .
\end{equation}
The fact that these functions are inverse to each other is expressed as
\begin{equation}\label{7}
    \Rt(R(\rt)) = \rt, \qquad R(\Rt(r)) = r .
\end{equation}

\section{Mirror Shapes} \label{sec:ms}

Let $\vec{I}$ and $\vec{R}$ denote the unit incidence and unit reflection
vectors at a mirror surface.  Let $\vec{N}$ denote a vector normal to a
point on a mirror surface.  At point $(x,y,h(x,y))$ on the first mirror, 
these vectors are 
\begin{equation}\label{8}
    \vec{I} = \left[ \begin{array}{c} 0 \\ 0 \\ 1 \end{array} \right], 
    \qquad
    \vec{N} = 
    \left[ \begin{array}{c} -h_x \\ -h_y \\ 1_{\phantom{x}} \end{array} \right],
    \qquad
    \vec{R} = \left[ \begin{array}{c}
	          \xt - x \\ \yt - y \\ \htild - h
              \end{array} \right]
	      \frac{1}{S(x,y,\xt,\yt)} ,
\end{equation}
where $h = h(x,y)$, $\htild = \htild(\xt,\yt)$, subscripts $x$ and $y$ denote
partial differentiation with respect to the indicated variable, i.e. 
\begin{equation} \label{8.5}
    h_x(x,y) = \frac{\partial h(x,y)}{\partial x},
\end{equation}
and
\begin{equation}\label{9}
    S(x,y,\xt,\yt) = \sqrt{(\xt-x)^2 + (\yt-y)^2 + (\htild-h)^2}
\end{equation}
denotes the distance from point $(x,y,h)$ on the first mirror 
to the corresponding point $(\xt,\yt,\htild)$ on the second mirror.
The shape of the first mirror is determined by requiring that 
$\vec{I}$, $\vec{N}$,
and $\vec{R}$ are coplanar and that angle of incidence equals angle of 
reflection.  In other words, it is required that 
$\vec{I}\times\vec{N} = -\vec{R}\times\vec{N}$.  Computing these cross
products, we get
\begin{equation}\label{10}
    \vec{I}\times\vec{N} =
    \left[ \begin{array}{r} h_y \\ -h_x \\ 0_{\phantom{x}} \end{array} \right]
    \qquad
    \vec{R}\times\vec{N} =
    \left[ \begin{array}{c} 
        \yt - y + h_y(\htild-h) \\
        x - \xt - h_x(\htild-h) \\
	h_x(\yt-y) - h_y(\xt-x) 
    \end{array} \right]
    \frac{1}{S} ,
\end{equation}
where $S$ is shorthand for $S(x,y,\xt,\yt)$.  Equating the first two
components, we can easily solve for $h_x$ and $h_y$:
\begin{equation}\label{11}
    h_x = \frac{x-\xt}{S+\htild-h}, \qquad
    h_y = \frac{y-\yt}{S+\htild-h}.
\end{equation}

Because the light after reflecting off from the second mirror should be
traveling in the $z$ direction once again, it follows from simple geometry
that
\begin{equation}\label{12}
    \htild_{\xt} = h_x \quad \text{ and } \quad \htild_{\yt} = h_y .
\end{equation}

As in our earlier paper \citet{TV03}, we have the following important lemma that
tells us that the optical path length for an on-axis source through the system
is constant.
\begin{lemma} \label{lemma1}
    $P_0 = S + \htild - h$ is a constant.
\end{lemma}
\begin{proof}
    The quantity of interest, $S + \htild - h$, is a shorthand for
    $S(x,y,\xt,\yt) + \htild(\xt,\yt) - h(x,y)$.  Although it appears that there
    are four independent variables, there are in fact only two since variables
    $x$ and $y$ can be considered to be functions of $\xt$ and $\yt$ (or vice
    versa).  Hence, we need only show that the derivatives with respect to
    $\xt$ and $\yt$ vanish.  Using \eqref{12}, it is easy to check that
    \begin{equation}\label{13}
        \frac{d}{d \xt} 
	\left( S(x,y,\xt,\yt) + \htild(\xt,\yt) - h(x,y) \right)
	=
	G(\xt,\yt) \left( 1 - \frac{\partial x}{\partial \xt} \right)
	-
	K(\xt,\yt) \frac{\partial y}{\partial \xt} ,
    \end{equation}
    where
    \begin{equation}\label{14}
        G(\xt,\yt) = \frac{\xt-x+h_x(\htild-h)}{S} + h_x
    \end{equation}
    and
    \begin{equation}\label{15}
        K(\xt,\yt) = \frac{\yt-y+h_y(\htild-h)}{S} + h_y .
    \end{equation}
    Putting the two right-hand terms over a common denominator ($S$) and then
    substituting the expressions in \eqref{11} for $h_x$ and $h_y$, it is easy
    to see that both $G$ and $K$ vanish.  Hence, the derivative with respect
    to $\xt$ vanishes.  That the derivative with respect to $\yt$ also
    vanishes is shown in precisely the same way.
\end{proof}

An important consequence of this Lemma is a simple decoupling of the
differential equations for the mirror shapes.
\begin{theorem} \label{thm1}
    The shape of the first mirror is determined by the following pair of
    ordinary differential equations:
    \begin{equation}\label{16}
        h_x = \frac{x-\xt}{P_0}, \qquad h_y = \frac{y-\yt}{P_0},
    \end{equation}
    where $\xt$ and $\yt$ are known functions of $x$ and $y$.
    Similarly, the shape of the second mirror is determined by:
    \begin{equation}\label{17}
        \htild_{\xt} = \frac{x-\xt}{P_0}, \qquad \htild_{\yt} = \frac{y-\yt}{P_0},
    \end{equation}
    where $x$ and $y$ are known functions of $\xt$ and $\yt$.
\end{theorem}
\begin{proof}
    The equations are an immediate consequence of Lemma \ref{lemma1},
    \eqref{11}, and \eqref{12}.  The ``known'' formulae relating $\xt$ and $\yt$
    to $x$ and $y$ are easy to compute from the functions $R$ and $\Rt$.
    For example,
    \begin{equation}\label{18}
        \xt = \rt \cos \theta + \delta = \frac{\rt}{r} r \cos \theta + \delta
	    = \frac{\Rt(r)}{r} x + \delta 
	    = \frac{\Rt(\sqrt{x^2+y^2})}{\sqrt{x^2+y^2}} x + \delta
	    .
    \end{equation}
    The other relations are derived in a similar manner.
\end{proof}

We have two approaches that can be followed in solving for the 
mirror shapes.  First, we can solve each of equations \eqref{16} and
\eqref{17} separately; an advantage of this, for the case when the differential 
equations must be solved numerically, is that we can choose the $x$ or 
$\xt$ coordinates independently, say with equal step sizes.  
Second, we can solve either \eqref{16} or \eqref{17}, 
analytically or numerically, 
then find the other shape algebraically, as follows.
From Lemma \ref{lemma1} we can solve for $S$ and then square it to see that
\begin{equation}\label{19}
    S^2 = \left( P_0 - \htild + h \right)^2 = P_0^2 - 2 P_0 (\htild - h) + (\htild-h)^2 .
\end{equation}
Substituting $S^2 = (\xt-x)^2 + (\yt-y)^2 + (\htild-h)^2$, we see that the
quadratic terms involving $\htild-h$ cancel and so we are left with a simple
calculation for the difference:
\begin{equation}\label{20}
    \htild - h = \frac{P_0}{2} - \frac{(\xt-x)^2 + (\yt-y)^2}{2 P_0} .
\end{equation}
Hence, if we already know either $\htild$ or $h$, we can use \eqref{20} to
compute the other one.
An advantage of this method, for the case of numerical solutions of 
the differential equations, is that the step size chosen for one 
mirror will automatically map to the corresponding step on the other 
mirror.  Another way to say this is that the ray intersections on 
the first mirror map to the intersections of the same rays on the 
second mirror, facilitating visualization of the ray paths.

Let us return now to polar coordinates.  Put
\begin{equation}\label{21}
    \Ht(\rt,\theta) = \htild(\xt,\yt) ,
\end{equation}
where $\xt = \rt \cos \theta + \disp$ and $\yt = \rt \sin \theta$.
Using \eqref{3} and \eqref{4}, we get the following differential equations 
for $\Ht$:
\begin{equation}\label{22}
    \Ht_{\rt} = \htild_{\xt} \cos \theta + \htild_{\yt} \sin \theta 
    	      = \frac{(x-\xt)\cos\theta + (y-\yt)\sin\theta}{P_0}
    	      = \frac{(r-\rt)-\disp\cos\theta}{P_0} 
\end{equation}
and
\begin{equation}\label{23}
    \Ht_{\theta} = -\htild_{\xt} \rt \sin \theta + \htild_{\yt} \rt \cos \theta 
    	         =  \frac{-(x-\xt)\rt\sin\theta + (y-\yt)\rt\cos\theta}{P_0}
    	         =  \frac{\disp\rt\sin\theta}{P_0} 
		 =  \frac{\disp \yt}{P_0}.
\end{equation}
We can integrate 
in the radial direction along $\rt$ (holding $\theta$ constant) to get
%
\begin{equation}\label{24}
    \Ht(\rt,\theta) = \Ht(0,0) + \int_0^{\rt} \frac{R(u)-u-\disp\cos\theta}{P_0} du
                    = \Ht(0,0) + \frac{1}{P_0}\int_0^{\rt} (R(u)-u)du
		                 - \frac{\disp \rt \cos\theta}{P_0} .
\end{equation}
Note that the last term ($\disp\rt\cos\theta/P_0$) corresponds to the ``tilt''
of the mirror to accomodate the off-axis alignment of the two mirrors (i.e.,
the fact that $\disp \ne 0$).  Furthermore, this tilt term is especially simple
when written in cartesian coordinates:
\begin{equation}\label{25}
    \frac{\disp \rt \cos\theta}{P_0} = \frac{\disp (\xt - \disp)}{P_0} .
\end{equation}

It will be useful to have a polar-coordinate version of $h(x,y)$ as well. 
Put
\begin{equation} \label{610}
   H(r,\theta) = h(x,y) .
\end{equation}
Then from equation \eqref{20} and \eqref{3}-\eqref{4} we find 
\begin{equation} \label{611}
   H(r,\theta) = \Ht(\rt,\theta) 
                - \frac{P_0}{2}
                - \frac{ (\rt-r)^2 + 2(\rt-r) \delta \cos(\theta)
                        + \delta^2 }{2 P_0} .
\end{equation}

\section{Mapping and Apodization Relationship} \label{sec:ea}

 In this section, we show how to relate the pupil mapping 
function $R(\rt)$ to a specified amplitude apodization function 
$A(\rt)$.  Here $A(\rt)$ is the geometric gain factor relating the 
electric field amplitude in the entrance pupil $E(r)$ to that in 
the exit pupil $\Et(\rt)$.  We assume that $A$ is real.  We have 
\begin{equation} \label{27a}
     \Et(\rt) = A(\rt) E(r)         
\end{equation}
where, as above, $r$ and $\rt$ are related by the specified geometric 
mapping (equations \eqref{5} and \eqref{6}).

We now invoke conservation of energy and require that the 
intensity of light in the entrance pupil $I(r)$ be related to that 
in the exit pupil $\It(\rt)$ by
\begin{equation} \label{27b}
  \It(\rt) \rt d\rt d\theta d\lambda = I(r) r dr d\theta d\lambda  
\end{equation}
where $\lambda$ is wavelength and intensity is in units of energy 
per unit time per unit area per wavelength interval.

We then have that $I = |E|^2$ and $\It = |\Et|^2$, in appropriate 
units.  Combining \eqref{27a} and \eqref{27b} we get 
\begin{equation} \label{27c}
     A^2(\rt) \rt d\rt = r dr      
\end{equation}
where $r = R(\rt)$.  
%
From this it follows that 
\begin{equation}\label{29}
    R(\rt) R'(\rt) = A^2(\rt) \rt .
\end{equation}
In other words, 
\begin{equation}\label{30}
    \left( R(\rt)^2 \right)' = 2 A^2(\rt) \rt 
\end{equation}
which can be integrated to yield
\begin{equation}\label{31}
    R(\rt) = \pm \sqrt{\int_0^{\rt} 2 A^2(s) s ds} .
\end{equation}

Equation \eqref{31} is the fundamental relation that connects the amplitude
apodization function $A$ to the ray-mapping function $R$.  In the remainder of
this paper, we will develop several results that follow from this relation:
In Section \ref{sec:ex}, we give some explicit examples of pupil mapping.
In Section \ref{sec:ext}, we discuss extensions to different geometries.
In Section \ref{sec:refract}, we discuss replacing the off-axis mirrors 
with on-axis lenses and derive the corresponding equations for 
their surfaces.
In Section \ref{sec:oa}, we return to pupil-mapping mirrors and 
explore the off-axis aberrations analytically.

We end this section by remarking that the relationship between the functions
$R'$ and $A$ given for the 1-D case in \cite{TV03} (equation (32)) was derived
incorrectly.  Fortunately, the correction is simple: just replace $A$ with $A^2$
in equation (32) and in all subsequent equations that depend on this one.

%
%

\section{Examples of 2-Mirror Pupil Mapping} \label{sec:ex}

\subsection{Constant Mapping Function} \label{sec51}
           
In this subsection, we choose a constant mapping function and explore 
its implications.  We will find that this simple case leads to a 
total of 6 different types of 2-mirror systems, depending on the 
value of the constant.  This family includes the familiar afocal 
Cassegrain, Gregorian, and periscope systems, plus variants on 
these.

Let us take the intensity mapping function to be
\begin{equation} \label{612}
   A^2(\rt) = \alpha^2
\end{equation}
so that the amplitude mapping function is 
\begin{equation} \label{613}
   A(\rt) = \alpha
\end{equation}
where we choose the positive root, but allow alpha to take on any 
real value, positive or negative.  From equation \eqref{31}, the ray-mapping 
function is then 
\begin{equation} \label{614}
   r = R(\rt) = \alpha \rt,
\end{equation}
where again we choose the positive root, with no loss of generality.
   
Inserting this into equations \eqref{24} and \eqref{611}, 
and requiring that the first 
mirror be centered at the origin, $H(0,0) = 0$, we find the following 
solutions for the first and second mirror surfaces, respectively:
\begin{eqnarray} 
   H(r,\theta) &=& (1-\frac{1}{\alpha})\frac{r^2}{2 P_0} 
                  - \frac{r \delta \cos\theta}{P_0} \\[0.2in]
   \Ht(\rt,\theta) &=& - (1-\alpha) \frac{r^2}{2 P_0} 
                            - \frac{\rt \delta \cos(\theta)}{P_0}
                            + \frac{P_0}{2} - \frac{\delta^2}{2P_0} 
\end{eqnarray}
                      
For $\alpha \ne 1$, these equations describe paraboloidal mirrors, formed 
by the rotation of a parabola about an axis parallel to the z axis. 
For $\alpha = 1$, they describe flat mirrors tilted about axes parallel 
to the $y$ axis, i.e., a simple periscope.

Setting first derivatives of $H$ and $\Ht$ to zero,
we find that the axes of the $H$ and $\Ht$ paraboloids are both centered at 
\begin{equation} \label{616}
   (x,y) = \left( \delta \frac{\alpha}{\alpha-1}, 0 \right).
\end{equation}
The height of the vertex of each paraboloid, either real or 
projected, is at 

\begin{equation} \label{617}
   H_{\mbox{\scriptsize min}} = -\frac{\delta^2 \alpha}{2P_0(\alpha-1)}
\end{equation}
and 
\begin{equation} \label{618}
   \Ht_{\mbox{\scriptsize max}} = \frac{P_0}{2} - \frac{\delta^2 \alpha}{2 P_0 (\alpha-1)}.
\end{equation}
The difference in heights is 
\begin{equation} \label{}
   \Ht_{\mbox{\scriptsize max}} - H_{\mbox{\scriptsize min}} = P_0/2
\end{equation}
as expected from the property of $P_0$ as the additional optical path 
imposed by the two mirrors.

The second derivatives of $H$ and $\Ht$ are the inverse of twice the 
paraxial focal lengths, i.e., $H_{rr} = 1/(2F)$ and 
$\Ht_{\rt\rt} = -1/(2\Ft)$, where the signs are chosen 
to show that the light is incident on the mirrors from above and 
below, respectively.  Hence, we find these paraxial focal lengths:
\begin{eqnarray} 
   F & = & \frac{\alpha P_0}{2(\alpha-1)} \\
   \Ft & = & -\frac{P_0}{2(\alpha-1)}.
\end{eqnarray}
The sum of these is $F + \Ft = P_0/2$, indicating that the 
parabolas have the same focal point (given that their vertices are 
separated by this amount too), as is expected for a system with 
parallel light incident on it, and exiting from it.

\begin{figure}
\begin{center}
\text{\includegraphics[width=4.0in]{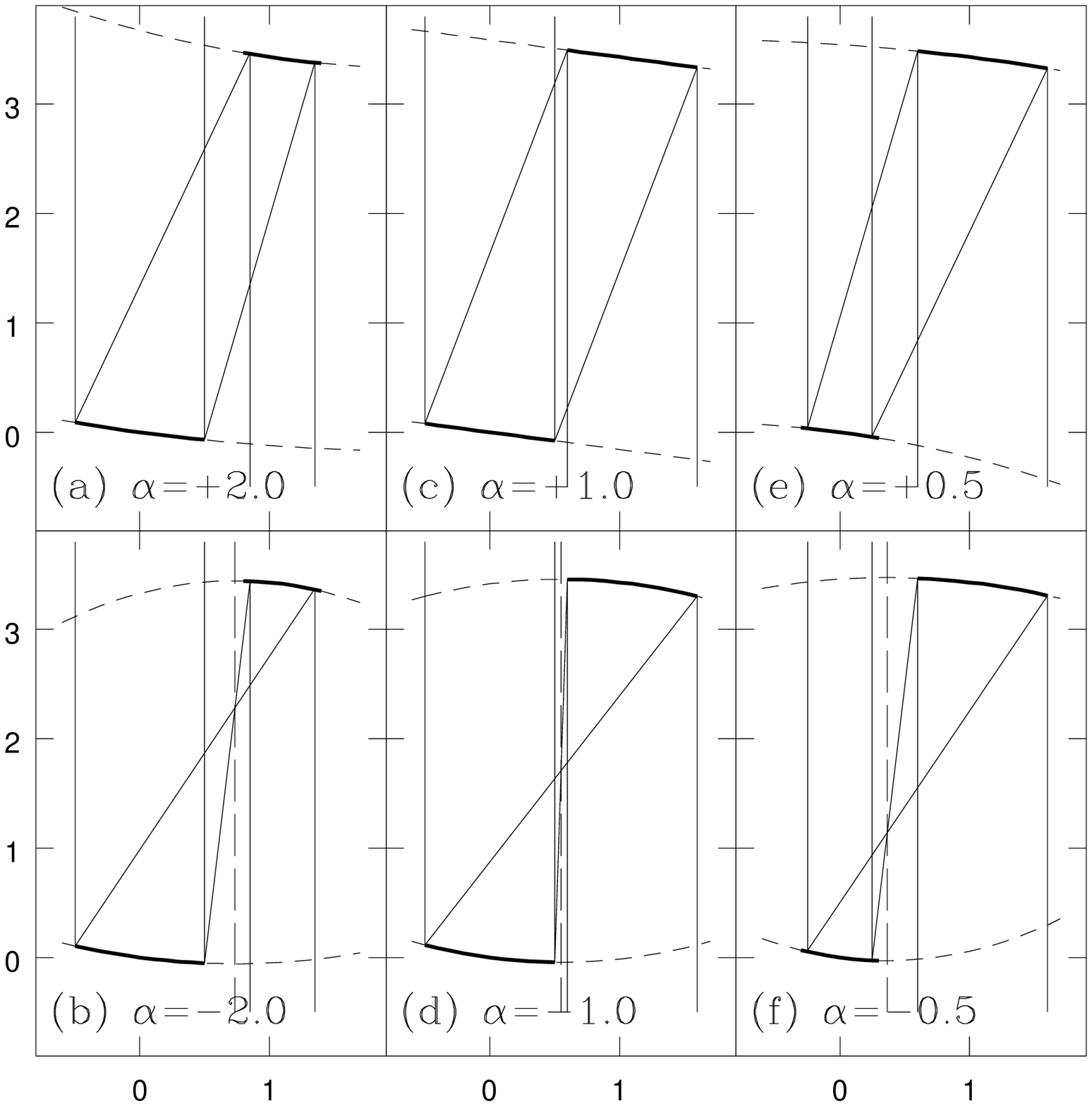}}
\end{center}
\caption{
This figure shows the family of 6 types of 2-mirror systems 
that can be obtained from the requirement that an input beam with a 
flat wavefront be mapped to an output beam also with a flat wavefront 
but with a uniform relative intensity of $\alpha^2$.  The 
short dashed lines show the large-scale surface of revolution of each 
paraboloid.  The heavy solid line shows the actual mirror surface. 
The long dashed line shows the common axis of revolution of the 
surfaces (outside the panel in the upper 3 panels).  In each case the 
vertices have the same vertical separation ($P_0/2 = 3.5$) and 
horizontal separation ($\delta = 1.1$), and the diameter of the 
larger mirror is $1$.  Input is from the upper left, and output is 
to the lower right.
(a) Cassegrain with $\alpha > 1$; confocal paraboloids, with common focus 
outside the panel (above and right).
(b) Gregorian with $\alpha < -1$; confocal paraboloids with common focus 
between the mirrors and on the axis of each mirror, as shown.
(c) Symmetric Cassegrain (periscope), $\alpha = 1$; two flat mirrors, 
common axis and focus at infinity.
(d) Symmetric Gregorian, $\alpha = -1$; confocal paraboloids, with focal 
points geometrically centered as shown.
(e) Inverted Cassegrain, $0 < \alpha < 1$; confocal paraboloids with 
common focus outside panel, exactly the same as (a) for reversed 
direction of light beam and $\alpha \rightarrow 1/\alpha$.
(f) Inverted Gregorian, $-1 < \alpha < 0$; confocal paraboloids with 
focus between the mirrors, same as (b) for reversed light and inverse 
$\alpha$.
}
\label{fig:1}
\end{figure}

These properties are summarized in Table \ref{tbl1}, and illustrated in Figure 
\ref{fig:1} which shows an $x-z$ cut through the mirrors and edge rays.

\subsection{Gaussian Mapping Function} \label{sec52}

In the context of searching for extrasolar planets, an ideal 
coronagraph would concentrate the incident starlight to a very 
compact image, free of the bright Airy rings that would otherwise be 
present.  The image-plane electric field is the Fourier transform of 
the pupil-plane electric field, and the Fourier 
transform of a Gaussian is a Gaussian.  Therefore, if we can generate 
a Gaussian amplitude distribution, the image from such a beam will 
have small sidelobes, within the approximation that the integral 
over a finite pupil is approximately the same as the integral over 
an infinite range.  (This latter restriction is removed in Sec. 5.3 
where the amplitude will be made equal to a prolate spheroidal
function.)  

For the Gaussian electric field case, we have the amplitude mapping 
function 
\begin{equation} \label{620}
   A(\rt) = c e^{-\frac{\rt^2}{2\sigma^2}}
\end{equation}
which generates the ray-mapping relation
\begin{equation} \label{621}
   R(\rt) = \sigma c \sqrt{\left( 1 - e^{-\rt^2/\sigma^2}\right)} .
\end{equation}
The constant $c$ is determined by the requirement that $R(\atwo) = \aone$ and
comes out to be
\begin{equation}\label{34}
  c = \frac{\aone}{\sigma \sqrt{\left( 1 - e^{-\atwo^2/\sigma^2}\right)}} .
\end{equation}
Hence,
\begin{equation}\label{503}
  R(\rt) = \aone
           \sqrt{\frac{1-e^{-\rt^2/\sigma^2}}{1-e^{-\atwo^2/\sigma^2}}} .
\end{equation}

\begin{figure}
\begin{center}
\text{\includegraphics[width=3.6in]{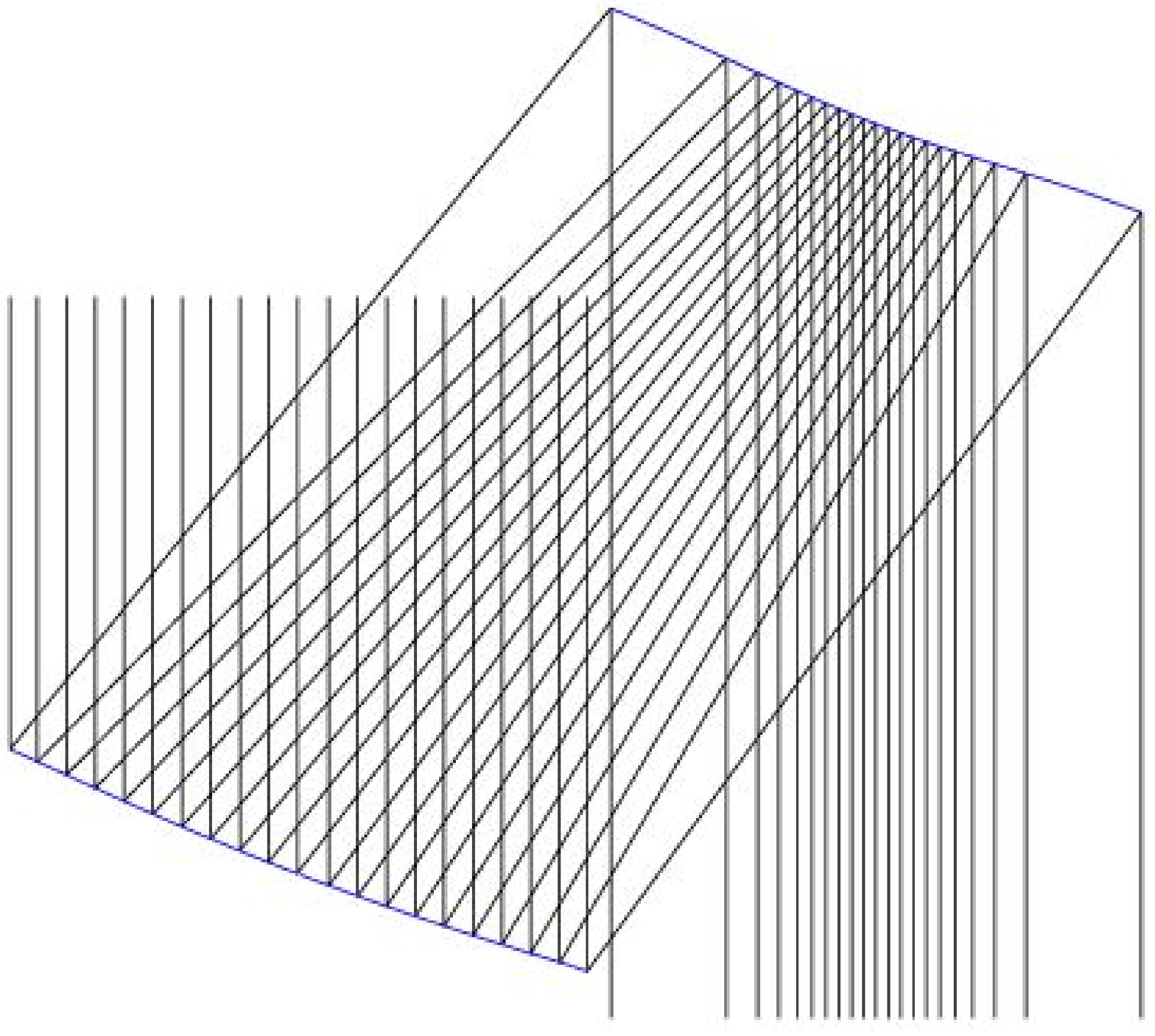}}
\text{\includegraphics[width=2.0in]{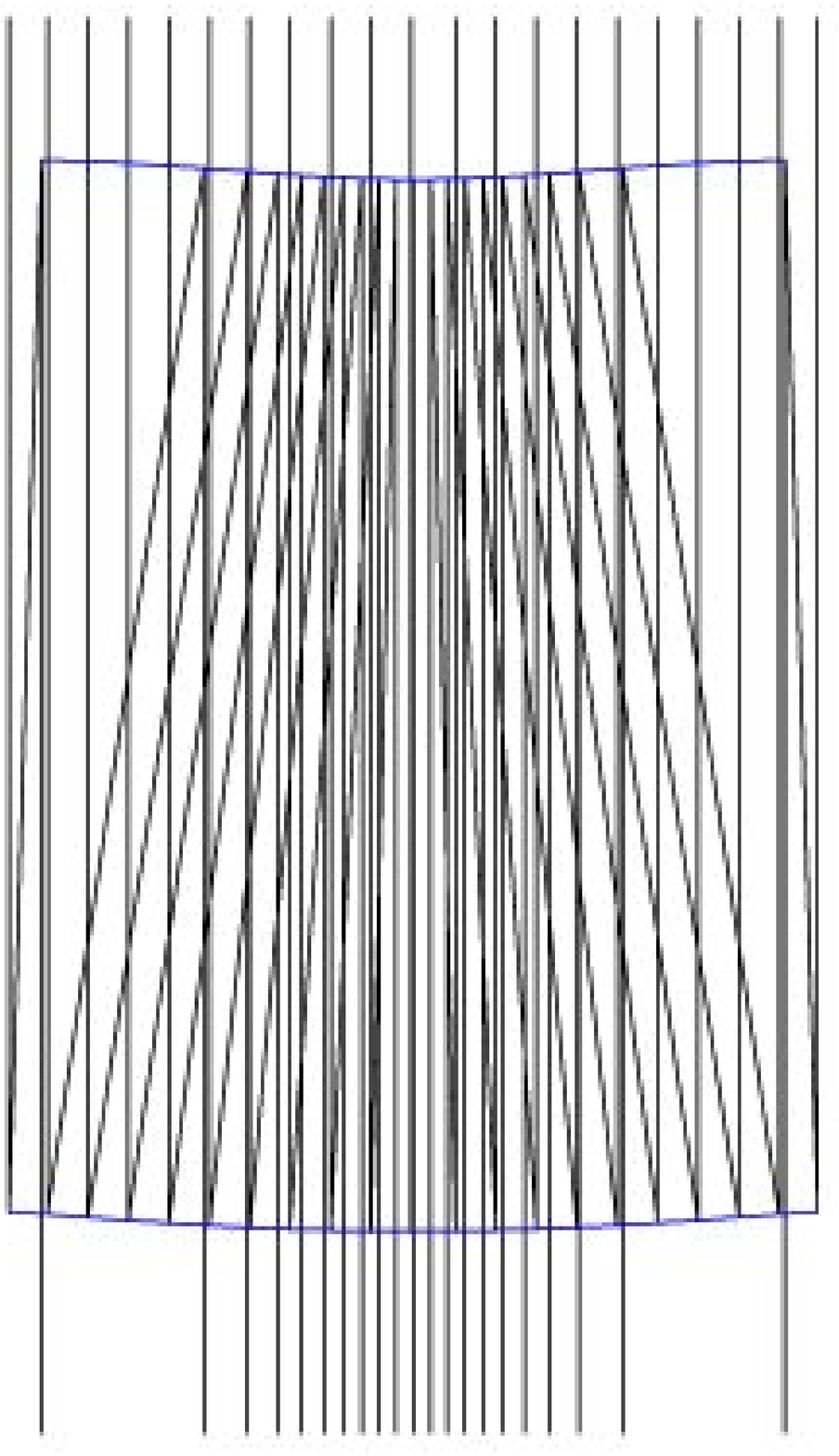}}
\end{center}
\caption{A Cassegrain design.
Parallel light rays come down from above, reflect off the bottom
mirror, bounce up to the top mirror, 
and then exit downward
as a parallel bundle with a concentration of rays in the center
of the bundle and thinning out toward the edges---that is, the
exit bundle is apodized, but with no loss of light.
}
\label{fig:2}
\end{figure}

\begin{figure}
\begin{center}
\text{\includegraphics[width=3.9in]{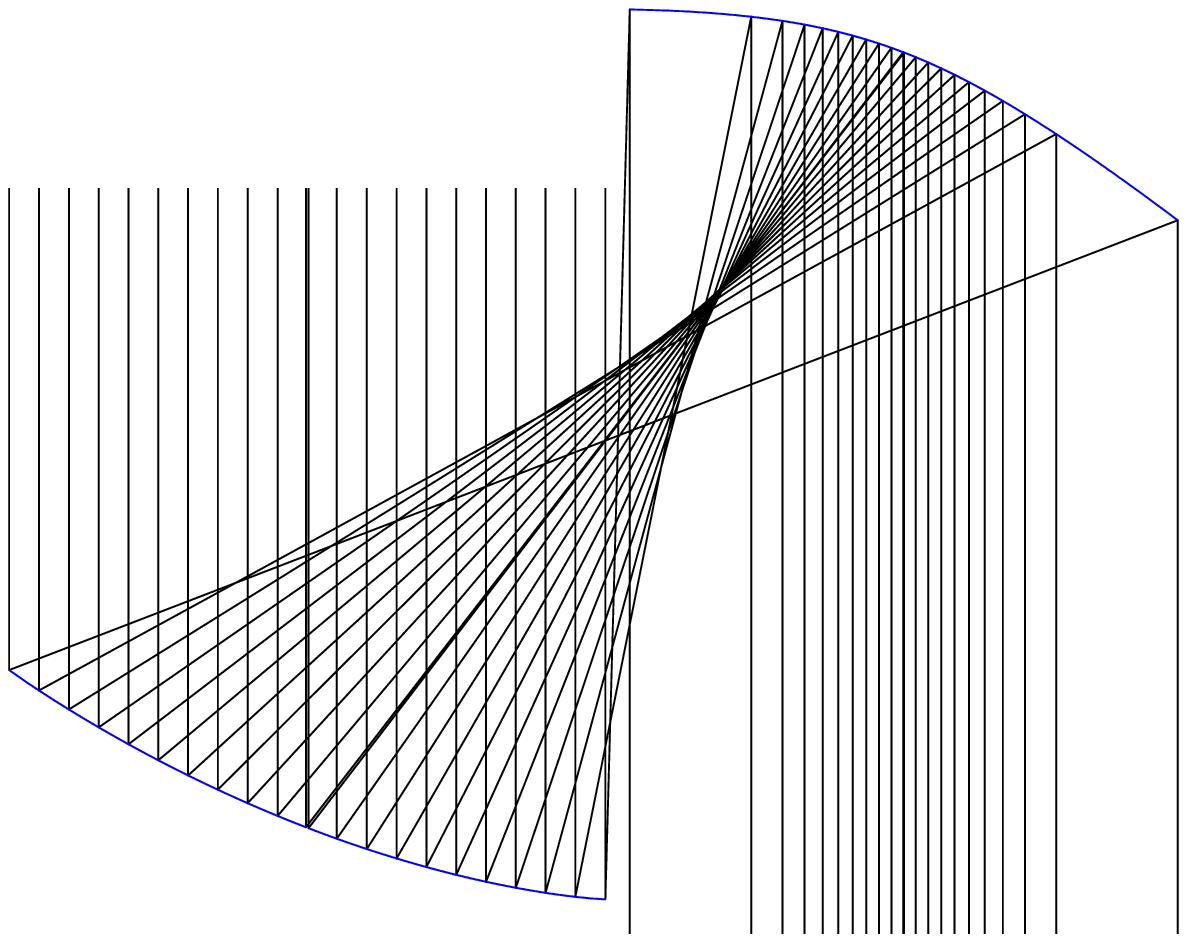}}
\text{\includegraphics[width=2.15in]{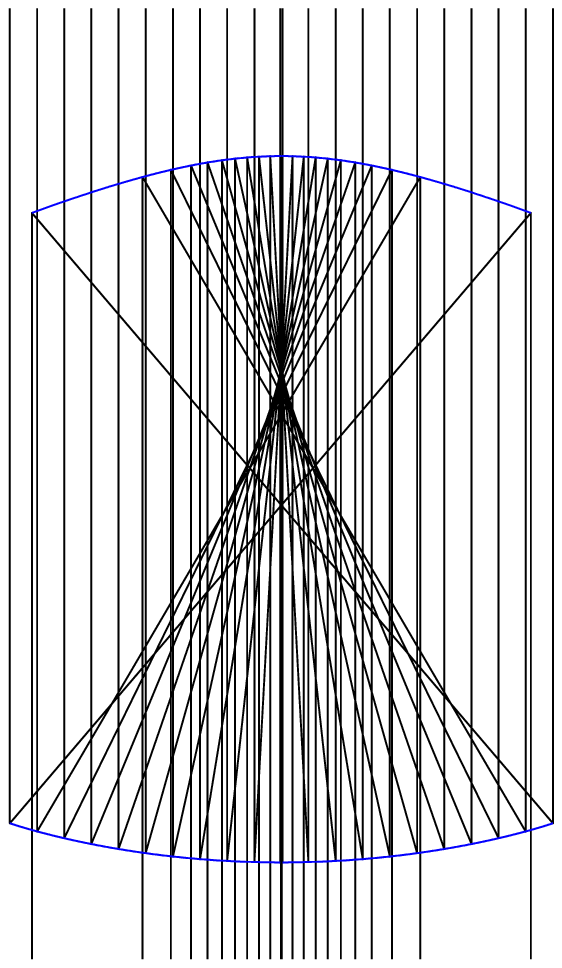}}
\end{center}
\caption{A Gregorian design.
Similar to Figure \ref{fig:2}, but since both mirrors are concave there is 
an intermediate focus, albeit a poor focus with a large halo.  
Nevertheless the rays emerge parallel, just as in Figure \ref{fig:2}.  An 
advantage here might be in manufacturability.
}
\label{fig:3}
\end{figure}

The equations for the surfaces $H$ and $\Ht$ can be solved 
numerically.  Typical results are shown in Figures \ref{fig:2} and \ref{fig:3}, 
where both the side view and end view are given, and selected rays are 
drawn.

\subsection{Prolate Spheroidal Wave Functions and Related Apodizations} 
\label{sec53}

On a finite interval, the prolate spheroidal wave function plays a role 
similar to that of a Gaussian on the infinite interval.  In 
particular, for both cases, the product of the width of the function 
and the width of its Fourier transform is minimal, compared to all 
other functional forms.  

\begin{figure}
\begin{center}
\text{\includegraphics[width=4.0in]{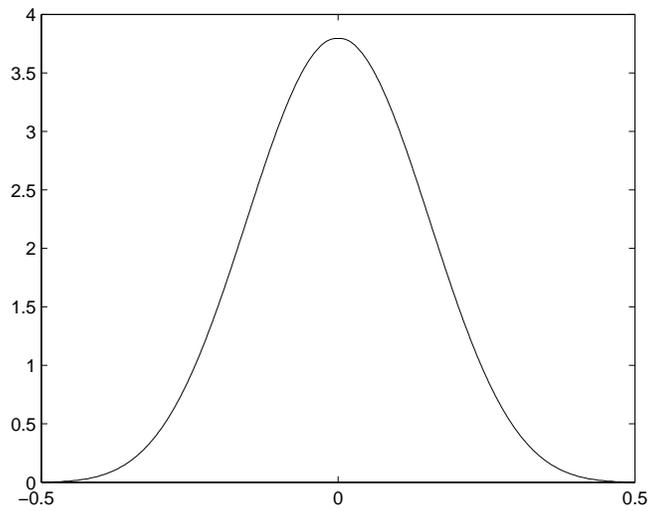}}
\end{center}
\caption{An energy conserving apodization providing contrast of $10^{-10}$ 
from $4 \lambda/D$ to $60 \lambda/D$.  As explained in Section 
\ref{sec:oa} (and in \cite{TV03}), the
unitless angle $4 \lambda/D$ only corresponds to a sky angle in the case where
the apodization function is identically one.  For non-trivial apodizations,
such as this one, the off-axis rays get magnified by a factor related to $A(r)$.
Hence, the intensity-weighted average magnification given by 
\eqref{704} (using \eqref{703}) evaluates to $2.16$ and
therefore $4 \lambda/D$ corresponds to $(4/2.16) \lambda/D = 1.85 \lambda/D$ 
as an angle on the sky.}
\label{fig:4}
\end{figure}

\begin{figure}
\begin{center}
\text{\includegraphics[width=4.0in]{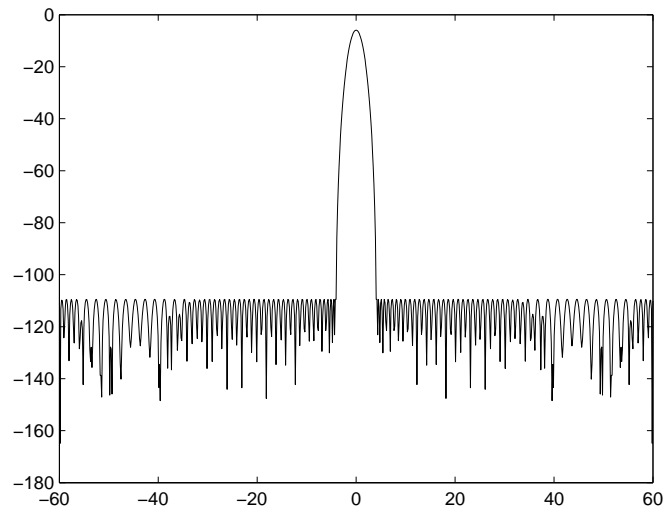}}
\end{center}
\caption{The on-axis point spread function for the apodization shown in Figure
\ref{fig:4}.  High contrast for an on-axis point source occurs at $4
\lambda/D$.  But, as explained in the previous caption, an off-axis source
such as a planet having, say, an angle of $2 \lambda/D$ in the sky will appear
mostly at $2 \times 2.16$ or $4.32 \lambda/D$ and is therefore detectable
in principle.}
\label{fig:5}
\end{figure}

Thus for a coronagraph, a likely optimal design is one in which the 
electric field across the pupil of an imaging lens is distributed as 
a prolate spheroidal wave function for 1-D optics and as a
generalized proloated spheroidal wave function for 2-D optics 
(see \cite{ref:Slepian}).  \citet{KVLS04} were the first to notice that
these functions play a fundamental role in shaped-pupil coronagraphs.
But, it turns out that one can get a slightly tighter inner working angle by
using an apodization function that is specifically designed to achieve the
desired contrast in the specified dark zone.  Optimization models designed in
this manner are described in \citet{VSK03,VSK02}.
We show in Figure \ref{fig:4} one such function.
In Figure \ref{fig:5} we 
show the point-spread function illustrating that the sidelobes are lower 
than $10^{-10}$ in intensity, nominally adequate for searching for 
Earth-like planets.

\section{Extensions} \label{sec:ext}

In this section, we briefly consider some generalizations to the basic set up
considered so far.

\subsection{Cassegrain vs. Gregorian Design} \label{sec:gk}

As we saw in \eqref{30}, the apodization function $A$ is related to the
square of the transfer function $R$ and, in \eqref{31}, there arose two choices
for the square root.   Suppose we choose the positive root.
Figure \ref{fig:2} shows two views of the resulting optical
system associated with the apodization function shown in Figure \ref{fig:4}.  
The first view is of the $(x,z)$-plane whereas the second view is of
the $(y,z)$-plane.  Since the secondary is convex, we
refer to this design as a {\em Cassegrain} 
design.

If we choose the negative square root, then we arrive at a different
optical system---one with a (rather poor) focal plane between the two mirrors.
This optical system is shown in Figure \ref{fig:3}.  We call this design a
{\em Gregorian} design, because the secondary is concave.  

The choice between a Cassegrain versus a Gregorian design is largely a question
of manufacturability; mathematically they perform the same.

%

\subsection{Concentric (On-Axis) Designs} \label{sec:conc}

When the second mirror is assumed to be of smaller aperture than the first, 
it is possible to consider an on-axis design.  In this case, $\Rt$ must map 
the annulus defined by the interval $[\atwo, \aone]$ of radii
bijectively onto $[0, \atwo]$.  All formulas in the previous sections 
remain unchanged.

\section{Refractive Elements} \label{sec:refract}

\begin{figure}
\begin{center}
\text{\includegraphics[width=4.0in]{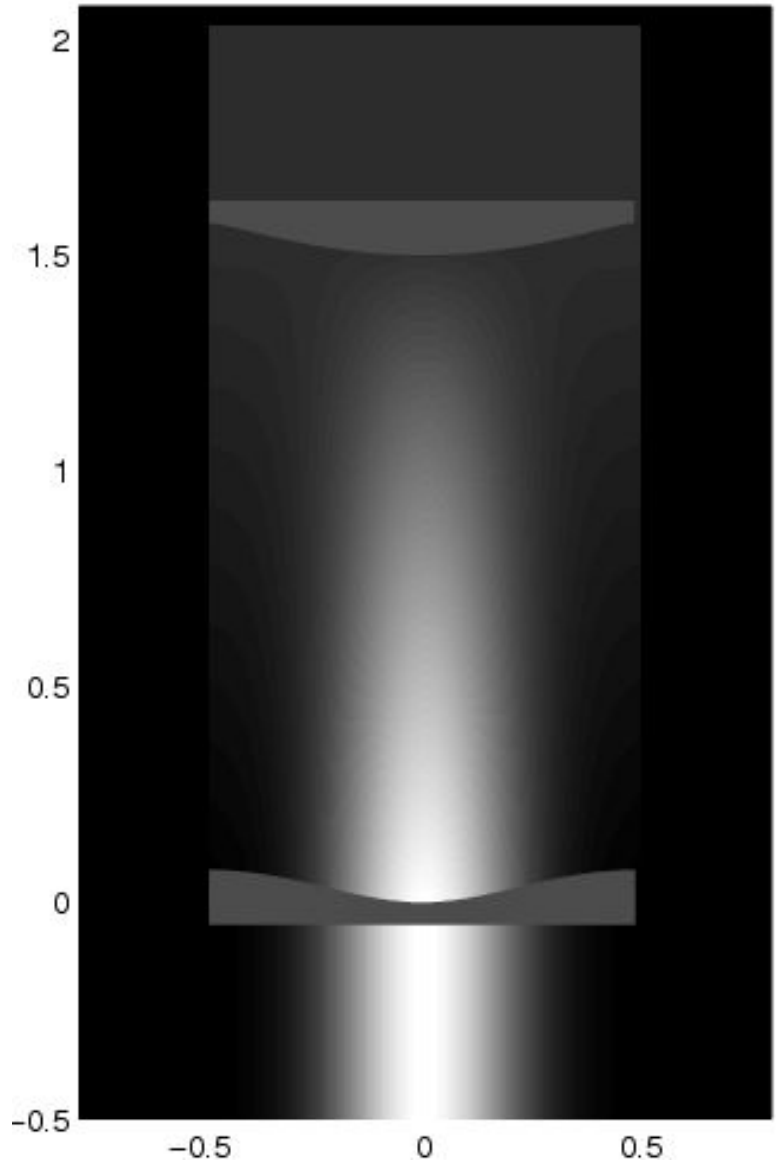}}
\end{center}
\caption{Pupil mapping via a pair of properly figured lenses.
This Galilean arrangement, with one convex and one concave lens, 
is the result of using $c>0$ in the Gaussian mapping function in Eqns. 
\eqref{620}--\eqref{34}.
If we had used $c<0$ the resulting lenses would both be 
convex, and the beam would have had a waist (an approximate focus) 
between the lenses.
}
\label{fig:6}
\end{figure}

We can replace mirrors $\Mone$ and $\Mtwo$ with coaxial lenses.
We require that $\Mone$ and $\Mtwo$ are plano on their outward-facing
surfaces, where the entering and exiting rays are parallel (see Figure 
\ref{fig:6}).
Assume that the lenses have refractive index $n$, which 
is constant over the desired band of wavelengths.  
It is easy to show that the mirror figures depend only on radius $r$.
Hence, mirror $\Mone$'s lower surface is defined by a function $h(r)$
and $\Mtwo$'s upper surface is given by a function $\htild(\rt)$.
Repeating the derivation at the beginning of section 
\ref{sec:ms} using the refractive form of Snell's law we 
derive the following analogue of equation \eqref{11}:
\begin{equation}\label{420}
    h_r(r) = \frac{r-\rt}{nS+\htild-h}.
\end{equation}
And, corresponding to equation \eqref{12}, we have
\begin{equation}\label{421}
    \htild_{\rt}(\rt) = h_r(r).
\end{equation}
Interestingly, Lemma \ref{lemma1} changes to:
\begin{lemma} \label{lemma4}
    $Q_0 = \frac{1}{n}S + \htild - h$ is a constant.
\end{lemma}
Hence, the analogue of Theorem \ref{thm1} is more complicated:
\begin{theorem} \label{thm2}
    The shape of the first lens is determined by the following differential 
    equation:
    \begin{equation}\label{430}
        h_r = \frac{r-\rt}{ \sqrt{n^2Q_0^2 + (n^2-1)(r-\rt)^2} }
    \end{equation}
    where $\rt$ is a known function of $r$.
    Similarly, the shape of the second lens is determined by:
    \begin{equation}\label{431}
        \htild_{\rt} = \frac{r-\rt}{ \sqrt{n^2Q_0^2 + (n^2-1)(r-\rt)^2} },
    \end{equation}
    where $r$ is a known function of $\rt$.
\end{theorem}
\begin{proof}
    From \eqref{420} and Lemma \ref{lemma4}, we see that
    \begin{equation}\label{600}
        h_r = \frac{r-\rt}{Q_0 + (n-1/n)S}
    \end{equation}
    Since $S^2 = (r-\rt)^2 + (h-\htild)^2$, we can write the invariant $Q_0$
    as
    \begin{equation} \label{601}
        Q_0 = \frac{1}{n} S + \sqrt{S^2 - (r-\rt)^2}.
    \end{equation}
    We can rearrange \eqref{601} into a quadratic expression in $S$ and then use
    the formula for the roots of a quadratic equation to express $S$ 
    in terms of $Q_0$ and $r-\rt$:
    \begin{equation} \label{602}
        S = \frac{-\frac{Q_0}{n} + \sqrt{Q_0^2 + (1-\frac{1}{n^2})(r-\rt)^2}}%
	         {1-\frac{1}{n^2}} .
    \end{equation}
    From this expression, we get that 
    \begin{equation} \label{603}
        Q_0 + ( n - 1/n )S = \sqrt{n^2Q_0^2 + (n^2-1)(r-\rt)^2} .
    \end{equation}
    Combining \eqref{600} with \eqref{603}, we get the result claimed.
\end{proof}

Interestingly, the equivalent of equations \eqref{430} and \eqref{431} 
was found by \citet{Kreuzerpatent},
in a patent application for a pair of lenses that 
could change the Gaussian-distributed beam from a laser into a 
sometimes more useful uniform-intensity beam.  Since light is 
reversible, Kreuzer's goal and ours are essentially the same.  Both 
before and after Kreuzer's discovery, there have been many papers 
pursuing similar ends;  as a recent example, we mention \citet{HJ03},
who designed and fabricated a pair of convex 
lenses for this purpose.  Our example, shown in Figure \ref{fig:6}, was derived 
using equations \eqref{620}--\eqref{503}, 
with a value $c > 0$, and gives one positive and 
one negative lens.  If we had used $c < 0$, we would have found both 
lenses to be positive, and this is effectively what Hoffnagle and 
Jefferson did in their experiment.

\section{Mapping Elliptical Pupils to Circular Ones} \label{sec:ellip}

The current baseline design for the Terrestrial Planet Finder space telescope
involves an $8 \times 3.5$m primary mirror and a square downstream deformable
mirror for wavefront control.  This disparity of shapes introduces the need to
reshape the pupil using anamorphic mirrors, which introduces the opportunity
also to develop unique pairs of anamorphic mirrors so as to apodize the exit
pupil as desired for high-contrast imaging.  We discuss this design problem
here.  So, we assume that mirror $\Mone$ is elliptical
having semimajor axes $a$ and $b$: $\{(x,y): (x/a)^2 + (y/b)^2 \le 1\}$.
We assume that $\Mtwo$ is a circular mirror of radius $\atwo$,
that $\Mone$ is uniformly illuminated, and that the beam leaving
mirror $\Mtwo$ is apodized according to a given function $A(\rt)$ that 
depends only on the radius $\rt$.  In this setup, the obvious hope would be
that the transfer function $\Rt$ from the circular case can simply be
stretched as needed at each angle $\theta$:
\begin{equation} \label{401}
    \rt = 
    \Rt\left(\frac{r\atwo}{ab}\sqrt{a^2 \sin^2\theta + b^2 \cos^2\theta}\right) 
\end{equation}
(the scaling of $r$ inside the function $\Rt$ is chosen so that the result
will be between $0$ and $\atwo$).
But, this transformation implies that the simple angular map 
$\thetat = \theta$ is no longer adequate; we need to introduce a
$\theta$-transfer function 
\begin{equation} \label{403}
    \theta = \Theta(\thetat)
\end{equation}
and its inverse
\begin{equation} \label{404}
    \thetat = \Thetat(\theta) .
\end{equation}
Given this, the inverse transformation for $\Rt$ is easy to find:
\begin{equation} \label{402}
    r = \frac{ab/\atwo}{\sqrt{a^2 \sin^2\Theta(\thetat)+b^2 \cos^2\Theta(\thetat)}}
        R(\rt) .
\end{equation}
From the usual change of variables, we get that
\begin{equation}\label{404a}
    r dr d\theta = 
        \frac{a^2b^2/\atwo^2}{a^2 \sin^2\Theta(\thetat) + b^2 \cos^2\Theta(\thetat)}
	    R(\rt) R'(\rt) \Theta'(\thetat) d\rt d\thetat 
\end{equation}
and our aim is to have
\begin{equation} \label{405}
    r dr d\theta = A(\rt)^2 \rt d\rt d\thetat .
\end{equation}
Combining \eqref{404a} and \eqref{405}, we see that 
\begin{equation} 
    A(\rt)^2 \rt = R(\rt) R'(\rt) \frac{\gamma}{\atwo^2} \label{406}
\end{equation}
\begin{equation} 
    \frac{a^2b^2}{a^2 \sin^2\Theta(\thetat) + b^2 \cos^2\Theta(\thetat)}
    \Theta'(\thetat) = \gamma .  \label{407}
\end{equation}
where $\gamma$ is a constant determined by a boundary condition
to be discussed shortly.
Integrating differential equation \eqref{407}, we get that
\begin{equation} \label{408}
    \thetat
    =
    \frac{ab}{\gamma}\tan^{-1}\left( \frac{a}{b} \tan \theta \right) .
\end{equation}
Since $\theta = \pi/2$ when $\thetat = \pi/2$, it follows that
\begin{equation} \label{409}
    \gamma = a b .
\end{equation}
Hence, 
\begin{equation} \label{410}
    \Thetat(\theta) = \tan^{-1}\left( \frac{a}{b} \tan \theta \right) ,
\end{equation}
\begin{equation} \label{411}
    \Theta(\thetat) = \tan^{-1}\left( \frac{b}{a} \tan \thetat \right) ,
\end{equation}
and $R(\rt)$ is determined by integrating
\begin{equation} \label{412}
    A(\rt)^2 \rt = R(\rt) R'(\rt) \frac{ab}{\atwo^2} .
\end{equation}
Converting polar functions $R(\rt)$ and $\Theta(\thetat)$ into cartesian 
equations for $x$ and $y$ as functions of $\xt$ and $\yt$, we can finally
calculate the mirror shapes using the differential equations in Theorem
\ref{thm1}.

\section{Off-Axis Performance} \label{sec:oa}

\begin{figure}
\begin{center}
\text{\includegraphics[width=4.0in]{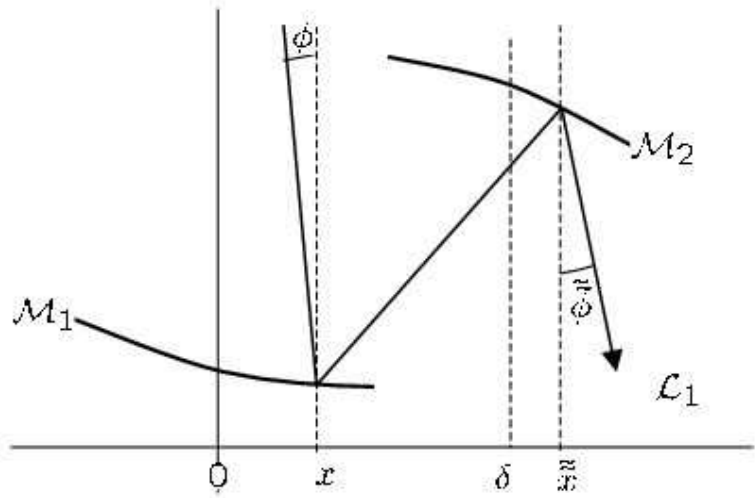}}
\end{center}
\caption{Ray-trace of an off-axis source as it passes through a two-mirror
pupil mapper.}
\label{fig:7}
\end{figure}

In this section we do a careful ray-trace analysis for an off-axis
source.  Suppose that the infinitely remote source lies in the $(x,z)$-plane
and is oriented at an angle $\phi$ from vertical so that its unit incidence
vector is $(-\sin \phi, 0, \cos \phi)$.  To keep the analysis manageable,
we will only trace rays in the $(x,z)$-plane.  
In polar coordinates, this is the $(r,z)$-plane and $\theta = 0$.
The incoming rays can be
parametrized by the $x$-coordinate at which they hit the first mirror.  Fix
an $x$ and consider such a ray.  A ray trace is shown in Figure
\ref{fig:7}.
Throughout this section, anytime a function
is a function of $y$ or $\yt$ (among other variables), we will suppress this
dependence since these variables are zero in this section.  Also, for
notational convenience we extend the definitions of functions $R$ and $\Rt$ to
negative values by making the functions odd:
\begin{equation} \label{202}
    R(-\rt) = - R(\rt) \qquad \text{and} \qquad \Rt(-r) = - \Rt(r).
\end{equation} 

Since the rays are entering the system at an angle, the place where the 
the reflection hits the second mirror is displaced, say by $\Dxt$, from the
point $\xt$ where an on-axis ray hit this second mirror.  
We begin with this displacement.
\begin{lemma} \label{lemma2}
    $\Dxt = S(x,\xt) \phi + o(\phi)$.
\end{lemma}
\begin{proof}
  This is exactly Lemma 1 in \cite{TV03}.
\end{proof}

A second quantity of importance is the angle $\phit$ at which a light ray
reflects off from the second mirror.  We are interested in how this angle
depends on the position $x$ and angle $\phi$ at which it hit the first mirror:
\begin{lemma} \label{lemma3}
    $\phit = \displaystyle \frac{\phi}{\Rt'(x)} + o(\phi)$.
\end{lemma}
\begin{proof}
  This is essentially Lemma 2 from \cite{TV03}.  More specifically,
  it follows directly from equations (15) and (16) in the proof of Lemma
  2 in \cite{TV03} that
  \begin{equation} \label{112}
    \phit \approx \frac{dx/d\xt}{S(x,\xt)} \dxt = \frac{dx}{d\xt} \phi .
  \end{equation}
  Then, $dx/d\xt(\xt) = R'(\xt) = 1/\Rt'(x)$.
\end{proof}

Lemmas \ref{lemma2} and \ref{lemma3} tell us that a ray incident on
the first mirror at $x$-coordinate $x$ and angle $\phi$ bounces off the second
mirror at $x$-coordinate $\xt(x) + S \phi = \Rt(x) + \delta + S \phi$ and
at angle $\phi/\Rt'(x)$.  We summarize this by writing
\begin{equation} \label{113}
  \left\{ x, \phi \right\}
  \stackrel{\Mone\rightarrow\Mtwo}{\longrightarrow}
  \left\{ \Rt(x) + \delta + S(x,\xt) \phi, \quad \frac{\phi}{\Rt'(x)} \right\} .
\end{equation}

Lemma \ref{lemma3} is both good and bad.  
It is good because, for apodizations
of interest, $\Rt'(x)$ is less than one for most $x$'s and hence there is a
built-in magnification---off-axis rays come out of the system at a steeper 
angle than they had on entry.  
It is bad because the rays are no longer
parallel and therefore cannot be focused to a diffraction-limited image.

We discuss in detail the issue of nonparallel rays, and how to remedy it,
in the next section.  We end this section with some further discussion of the
magnification effect.  It turns out to be most convenient to consider $\phit$
parametrized over $\rt$ rather than over $r$.  That is, we are interested in 
\begin{eqnarray} 
    \phit(\rt) &\approx& \frac{\phi}{\Rt'(R(\rt))} \label{701} \\
    		&=& R'(\rt) \phi \label{702} \\
		&=& \frac{A(\rt)^2 \rt \phi}{\sqrt{\int_0^{\rt} 2A(s)^2 s ds}} 
		    \label{703}.
\end{eqnarray}
Here, the last equality follows from \eqref{31}.  To arrive at an average
magnification, we take an intensity-weighted average of $\phit(\rt)/\phi$:
\begin{equation} \label{704}
    \mbox{magnif} 
    = \frac{\int_0^{\atwo} \phit(\rt) A(\rt)^2 \rt d\rt}%
           {\phi \int_0^{\atwo} A(\rt)^2 \rt d\rt} .
\end{equation}

\subsection{Pupil Restoration by System Reversal} \label{sec:roa}

\begin{figure}
\begin{center}
\text{\includegraphics[width=4.0in]{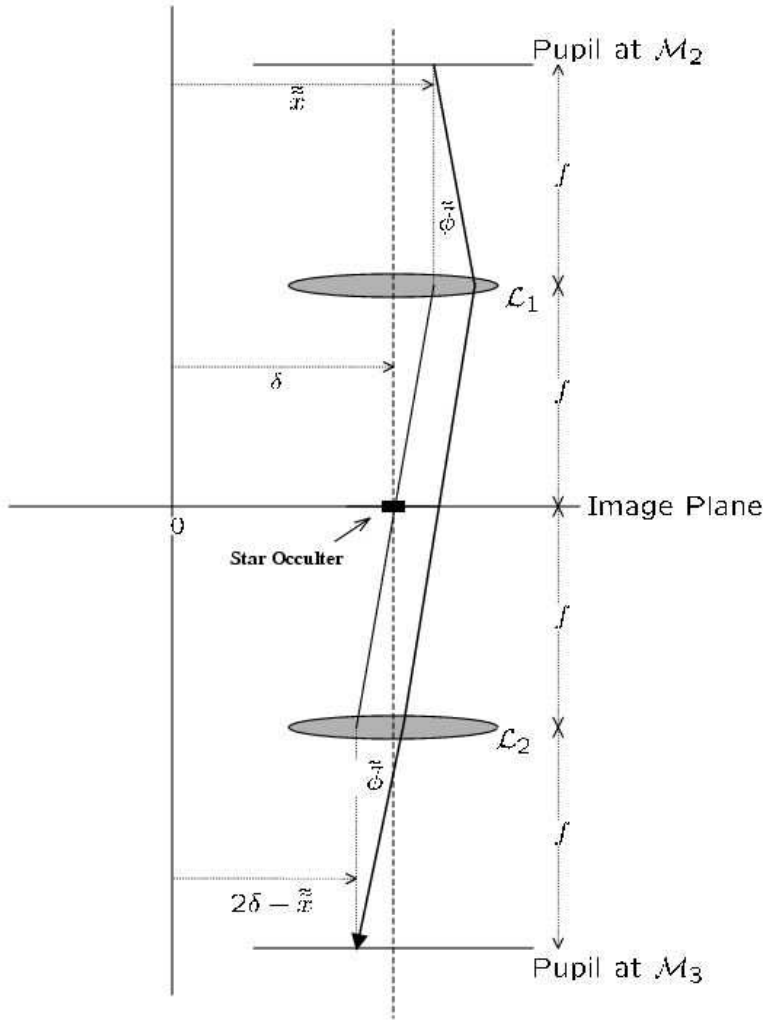}}
\end{center}
\caption{Ray-trace of an off-axis source from $\Mtwo$ to $\Mthree$ for a
4-mirror system. 
Note that by placing the lenses halfway between their corresponding mirrors
and the image plane and choosing their focal lengths to be this distance, we
get that the image of mirror $\Mtwo$ is exactly at $\Mthree$.  Hence,
each ray maps to a position reflected through the optical axis and comes out
with the opposite direction.}
\label{fig:8}
\end{figure}

\citet{Guy03} addresses this off-axis defocus question.  He recommends
placing a focusing element ${\cal L}_1$ 
after $\Mtwo$ followed by a star occulter
in the image plane, a beam-recollimating element ${\cal L}_2$, 
and finally a pupil mapping
system identical to the first two mirrors but set up exactly backwards.
Let's refer to these last two mirrors as $\Mthree$ and $\Mfour$.
If the focusing and recollimating elements (${\cal L}_1$ and ${\cal L}_2$)
are assumed to be ideal lenses having a common focal length, say $f$, and
if the distances from $\Mtwo$ to ${\cal L}_1$
and from ${\cal L}_2$ to $\Mthree$ are both also chosen to be $f$,
then the pupil at mirror $\Mthree$ is a reimaging of the pupil at 
$\Mtwo$ and so a ray hitting $\Mtwo$ at, say, position $\xtt$ and angle
$\phitt$ will hit $\Mthree$ at position $2\delta -\xtt$ and angle $-\phitt$.
We summarize this as
\begin{equation} \label{201}
  \left\{ \xtt, \phitt \right\}
  \stackrel{\Mtwo\rightarrow\Mthree}{\longrightarrow}
  \left\{ 2\delta -\xtt, -\phitt \right\}
\end{equation}
(see Figure \ref{fig:8}).

\begin{figure}
\begin{center}
\text{\includegraphics[width=3.5in]{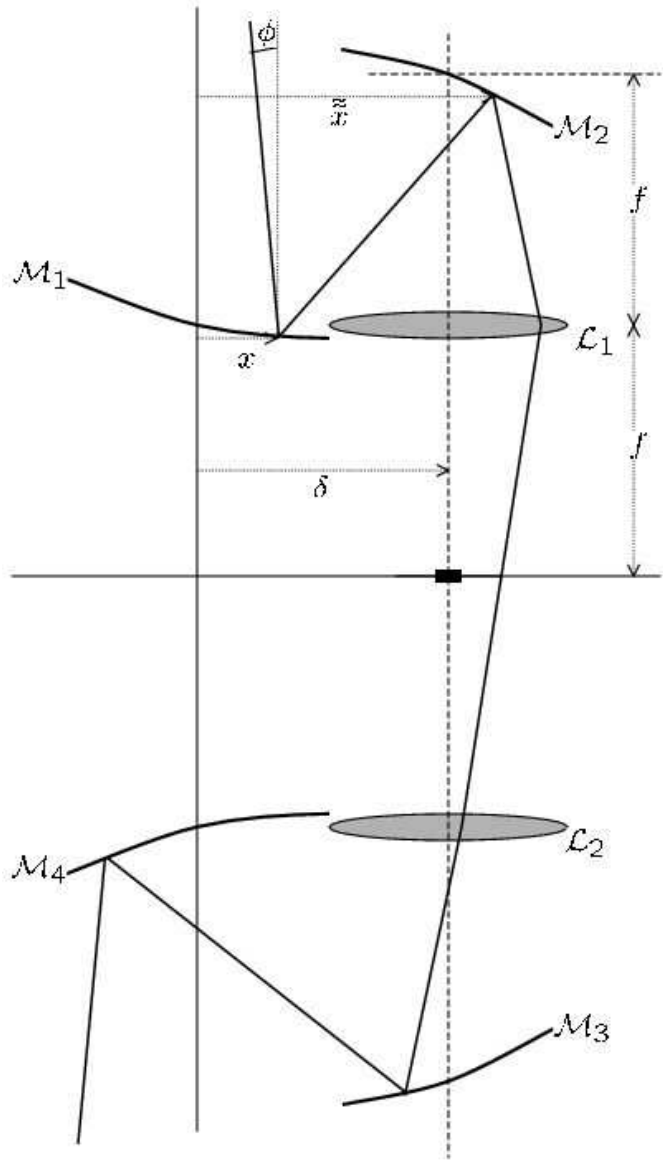}}
\end{center}
\caption{Ray-trace for the full 4-mirror system: 
two-mirror pupil mapper, lens system, and reversed two-mirror pupil mapper.  An
occulting spot is shown at the center of the image plane to block
on-axis starlight.}
\label{fig:9}
\end{figure}

Figure \ref{fig:9} shows a version of the full system in which the reversed
system folds back to the left.
The following theorem describes how an off-axis ray propagates from mirror
$\Mthree$ to $\Mfour$.

\begin{theorem} \label{thm3}
For the system shown in Figure \ref{fig:9}, a ray propagates from mirror
$\Mthree$ to mirror $\Mfour$ as follows:
\begin{equation} \label{114}
  \left\{ \xtt, \phitt \right\}
  \stackrel{\Mthree\rightarrow\Mfour}{\longrightarrow}
  \left\{ R(\xtt-\delta) + S(\xb,\xtt) \phitt, 
         \quad \frac{\phitt}{R'(\xtt-\delta)} \right\} 
\end{equation}
where $\xb$ denotes the point on mirror $\Mfour$ corresponding to an on-axis
ray impinging on mirror $\Mthree$ at $\xtt$.
\end{theorem}
\begin{proof}
Because the reversed pupil mapping system ($\Mthree$, $\Mfour$) 
is identical to the first one, we start by inverting the operation
given by \eqref{113}.  This inversion describes a ray propagating backwards
through mirrors $\Mtwo$ and $\Mone$.

From \eqref{113}, we have that 
\begin{equation} \label{330}
    \phitt = \frac{\phi}{\Rt'(x)}.
\end{equation}
Hence,
\begin{equation} \label{331}
    \phi = \Rt'(x) \phitt.
\end{equation}
We need to express $\Rt'(x)$ in terms of $\xtt$ (and perhaps $\phitt$).
To this end, we use equations \eqref{4}, \eqref{5}, and \eqref{6} to write
\begin{equation} \label{332}
    \xt = \rt+\delta = \Rt(r)+\delta = \Rt(R(\rt))+\delta 
        = \Rt(R(\xt-\delta))+\delta .
\end{equation}
Differentiating with respect to $\xt$, we get
\begin{equation} \label{333}
    1 = \Rt'(R(\xt-\delta)) R'(\xt-\delta) .
\end{equation}
Hence,
\begin{equation} \label{334}
    \Rt'(x) R'(\xt-\delta) = 1
\end{equation}
and so we get that
\begin{equation} \label{335}
    \phi = \frac{\phitt}{R'(\xt-\delta)}.
\end{equation}
Now, we remind the reader that our propagations
represent just the constant and linear terms in an expansion in the small
angular parameters $\phi$ and $\phitt$.  Since these angles are small, it
follows that $\xtt - \xt$ is also small.  We retain terms that are linear in
these small parameters but we drop higher order terms.  Hence, since the
right-hand side in \eqref{335} already is small, we can simply replace
$R'(\xt-\delta)$ with $R'(\xtt-\delta)$.

From \eqref{113}, we also have that
\begin{equation} \label{336}
    \xtt = \Rt(x) + \delta + S(x,\xt)\phi .
\end{equation}
We need to solve this for $x$.  From \eqref{7}, we see that
\begin{eqnarray} 
    x & = & R(\xtt - \delta - S(x,\xt)\phi) \label{337} \\
      & \approx & R(\xtt-\delta) - R'(\xt-\delta)S(x,\xt)\phi \label{3375} \\
      & = & R(\xtt-\delta) - S(x,\xt) \phitt \label{338} \\
      & \approx & R(\xtt-\delta) - S(\xb,\xtt) \phitt , \label{339}
\end{eqnarray}
where $\xb$ denotes the point on mirror $\Mone$ corresponding to an on-axis
ray impinging on mirror $\Mtwo$ at $\xtt$.

From \eqref{335} and \eqref{339}, we see that backwards propagation through
the $\Mone$-$\Mtwo$ system is given by
\begin{equation} \label{340}
  \left\{ \xtt, \phitt \right\}
  \stackrel{\Mtwo\rightarrow\Mone}{\longrightarrow}
  \left\{ R(\xtt-\delta) - S(\xb,\xtt) \phitt, 
         \quad \frac{\phitt}{R'(\xtt-\delta)} \right\} .
\end{equation}
To describe propagation through the $\Mthree$-$\Mfour$ system, we have to flip
this system about a horizontal axis, apply the above transformation, and then
flip back.  The flip operation leaves horizontal coordinates unchanged but
negates angles.  Hence, we get
\begin{eqnarray} 
  \left\{ \xtt, \phitt \right\}
  & \longrightarrow & \left\{ \xtt, -\phitt \right\} \label{341} \\
  & \longrightarrow & \left\{ R(\xtt-\delta) + S(\xb,\xtt) \phitt, 
		 \quad -\frac{\phitt}{R'(\xtt-\delta)} \right\} \label{342} \\
  & \longrightarrow & \left\{ R(\xtt-\delta) + S(\xb,\xtt) \phitt, 
		 \quad \frac{\phitt}{R'(\xtt-\delta)} \right\} \label{343}.
\end{eqnarray}
This completes the proof.
\end{proof}
%

We are now ready to combine the above ray propagation results to describe
propagation through the entire system:
\begin{theorem} \label{thm4}
  For the system shown in Figure \ref{fig:9},
  a ray entering the system at position $x$ and angle $\phi$, exits the system
  at position 
  $ - x - 2 \displaystyle\frac{S(x,\xt)}{\Rt'(x)}\phi $
  and at angle 
  $ \displaystyle\frac{-\phi}{1 - \frac{\Rt''(x)}{\Rt'(x)^2} S(x,\xt)\phi} $.
\end{theorem}
\begin{proof}
We can compose the maps given by \eqref{113}, \eqref{201}, and \eqref{114} 
to get that
\begin{eqnarray} 
  \left\{ x, \phi \right\}
  & \longrightarrow &  \label{115}
  \left\{ \Rt(x) + \delta + S(x,\xt) \phi
          , \quad \frac{\phi}{\Rt'(x)} \right\} \\
  & \longrightarrow &  \label{203}
  \left\{ -\Rt(x) + \delta - S(x,\xt) \phi
          , \quad -\frac{\phi}{\Rt'(x)} \right\} \\
  & \longrightarrow &  \label{120}
  \left\{ R\left(-\Rt(x) - S(x,\xt) \phi\right) - \frac{S(x,\xt)\phi}{\Rt'(x)} 
	  , \quad -\frac{\phi}{\Rt'(x)R'\left( -\Rt(x) - S(x,\xt) \phi\right)} 
  \right\} .
\end{eqnarray}
Since $R(\Rt(x)) = x$, we can differentiate this identity to derive simple
identities relating the derivatives of $R(\Rt(x))$ to the derivatives of
$\Rt(x)$.  From these identities, we easily see that
\begin{eqnarray} 
  R\left(\Rt(x) + S(x,\xt) \phi\right) 
  & = &  \label{116}
  R(\Rt(x)) + R'( \Rt(x) ) S(x,\xt)\phi + o(\phi)  \\
  & = &  \label{117}
  x + \frac{S(x,\xt)}{\Rt'(x)}\phi + o(\phi) 
\end{eqnarray}
and
\begin{eqnarray}
  R'\left( \Rt(x) + S(x,\xt) \phi\right)
  & = & \label{118}
  R'(\Rt(x)) + R''(\Rt(x))S(x,\xt)\phi + o(\phi) \\
  & = & \label{119}
  \frac{1}{\Rt'(x)} - \frac{\Rt''(x)}{\Rt'(x)^3} S(x,\xt)\phi + o(\phi) 
\end{eqnarray}
Using the fact that $R$ is an odd function, and therefore that $R'$ is even,
we substitute \eqref{117} and \eqref{119} into \eqref{120} to get the 
results claimed in the theorem.
\end{proof}

A comparison of the exiting ray angles from the 4-mirror corrected 
system (Theorem \ref{thm4}) compared to those from the 2-mirror uncorrected 
system (Lemma \ref{lemma4}) shows that the relative scatter of ray angles will be 
smaller in the corrected system by a factor which depends on the 
details of the function $\Rt$, but which in general will be roughly 
a factor of a few times the off-axis angle $\phi$; for typical 
planet-searching angles of a few arcseconds, this ratio could easily 
be less than $10^{-3}$, which is a significant reduction in scatter of ray 
angles.  (Specific estimates will be given in a future paper with 
numerical results.)

\begin{figure}
\begin{center}
\text{\includegraphics[width=4.0in]{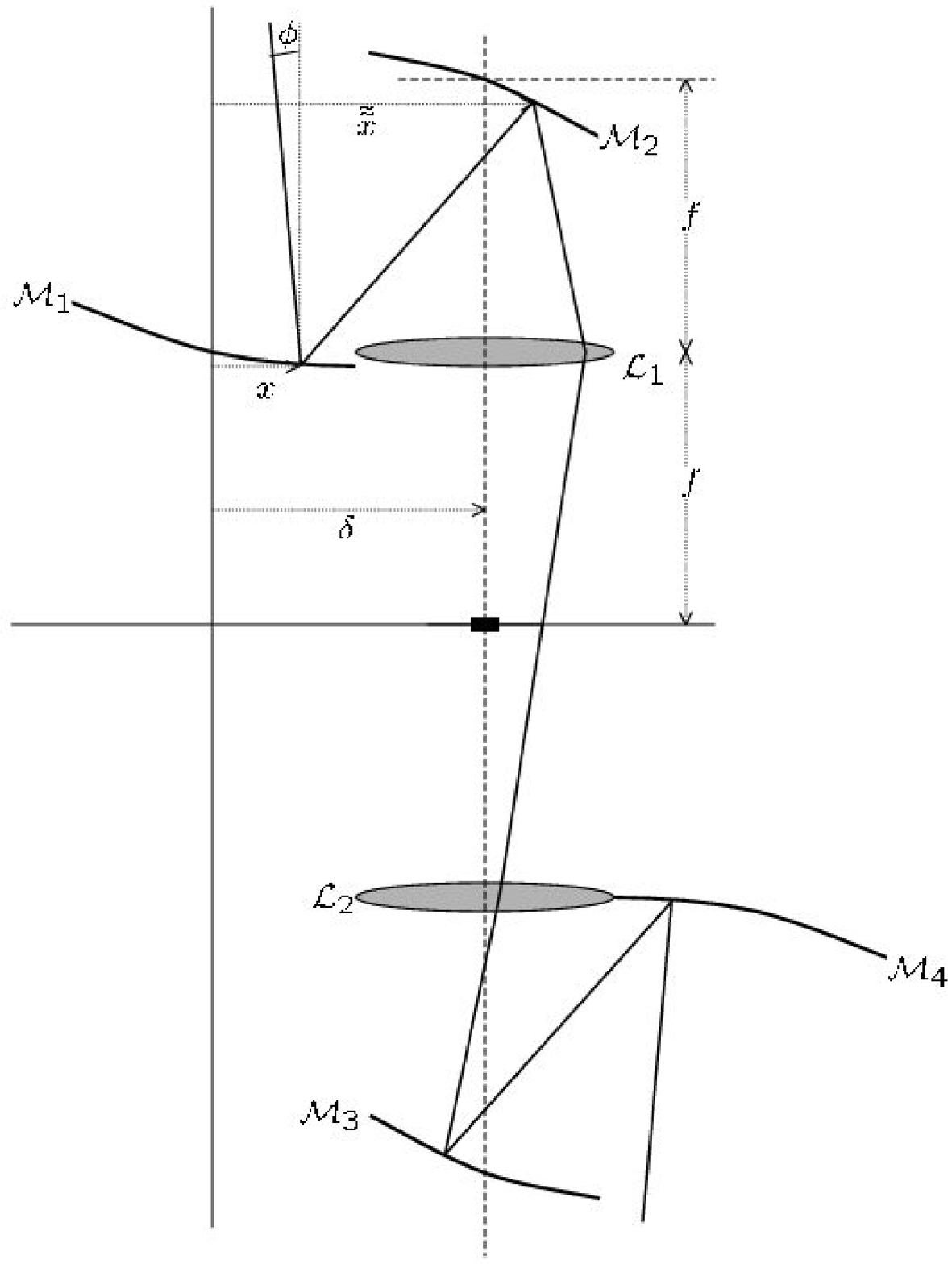}}
\end{center}
\caption{Ray-trace for the full 4-mirror system in which the reversed system 
continues to the right rather than folding back to the left.}
\label{fig:10}
\end{figure}

Figure \ref{fig:10} shows a version of the full system in which the reversed
pupil mapping folds off to the right.
This case is just like the previous one except that there is effectively a
symmetry reflection about the vertical axis $x = \delta$ that must be applied
before entering and after leaving the second system.  Flipping about this
vertical axis has the effect given by \eqref{201}.

\begin{theorem} \label{thm5}
  For the system shown in Figure \ref{fig:10},
  a ray entering the system at position $x$ and angle $\phi$, exits the system
  at position 
  $ 2 \delta - x - 2 \displaystyle\frac{S(x,\xt)}{\Rt'(x)}\phi $
  and at angle 
  $ \displaystyle\frac{-\phi}{1 - \frac{\Rt''(x)}{\Rt'(x)^2} S(x,\xt)\phi} $.
\end{theorem}
\begin{proof}
We need to compose the maps given by \eqref{113}, \eqref{201}, \eqref{201},
\eqref{114}, and \eqref{201}.  Of course, applying \eqref{201} twice in a row
simply undoes the effect and so we can simply form the composition of
\eqref{113}, \eqref{114}, and \eqref{201}:
\begin{eqnarray} 
  \left\{ x, \phi \right\}
  & \longrightarrow &  \label{315}
  \left\{ \Rt(x) + \delta + S(x,\xt) \phi
          , \quad \frac{\phi}{\Rt'(x)} \right\} \\
  & \longrightarrow &  \label{320}
  \left\{ R\left(\Rt(x) + S(x,\xt) \phi\right) + \frac{S(x,\xt)\phi}{\Rt'(x)} 
	  , \quad \frac{\phi}{\Rt'(x)R'\left( \Rt(x) + S(x,\xt) \phi\right)} 
	  \right\} \\
  & = & \label{321}
  \left\{ x + 2 \displaystyle\frac{S(x,\xt)}{\Rt'(x)}\phi, \quad
	  \frac{\phi}{1-\frac{\Rt''(x)}{\Rt'(x)^2}S(x,\xt)\phi} \right\} \\
  & \longrightarrow & \label{322}
  \left\{ 2 \delta - x - 2 \displaystyle\frac{S(x,\xt)}{\Rt'(x)}\phi, \quad
	  -\frac{\phi}{1-\frac{\Rt''(x)}{\Rt'(x)^2}S(x,\xt)\phi} \right\}
	  .
\end{eqnarray}
\end{proof}
Since the estimated aberrations $x$ and $\phi$ are the same in 
Theorems \ref{thm4} and \ref{thm5} 
(cf. Figures \ref{fig:9} and \ref{fig:10}), we see that these correcting 
schemes are equivalent in terms of their ability to reduce 
aberrations.

Finally, we remark that our analysis has ignored the beam walk that would be
introduced by the fact that mirrors $\Mtwo$ and $\Mthree$ are not
everywhere a distance $f$ from the corresponding lens.

\section{Summary} \label{sec:sum}
We derived equations for the shapes of mirrors and lenses capable 
of converting a uniform-intensity beam into a shaped-intensity beam 
(the pupil-mapping process).  We gave analytical estimates of the 
aberrations of a 2-mirror system, and the improved case of an 
aberration-corrected 4-mirror system.  We applied the results to 
several examples, including the familiar Cassegrain-Gregorian 
telescope designs, as well as beam shapes given by Gaussian and 
prolate spheroidal functions.   The general equations will allow 
the design of many types of optical systems, but in particular 
should be helpful in designing optics for telescopic searches for 
extrasolar planets.

{\bf Acknowledgements.}
This research was partially performed for the 
Jet Propulsion Laboratory, California Institute of Technology, 
sponsored by the National Aeronautics and Space Administration as part of
the TPF architecture studies and also under JPL subcontract number 1260535.
The first author received support from the NSF (CCR-0098040) and
the ONR (N00014-98-1-0036).

\bibliography{../lib/refs}   
\bibliographystyle{plainnat}   

\clearpage

\clearpage

\begin{table}
\centering
\begin{tabular}{|r|rrrrrr|} \tableline
Quantity (units)& 
\multicolumn{1}{c|}{$\alpha < -1$} &
\multicolumn{1}{c|}{$\alpha = -1$} &
\multicolumn{1}{c|}{$-1 < \alpha < 0$} &
\multicolumn{1}{c|}{$0 < \alpha < 1$} &
\multicolumn{1}{c|}{$\alpha = 1$} &
\multicolumn{1}{c|}{$\alpha > 1$} 
\\ \tableline

$x_{\mbox{\scriptsize min}} = \xt_{\mbox{\scriptsize max}}$ ($\delta$)& 
$1$ to $1/2$ & 
$1/2$ & 
$1/2$ to $0$ &  
$0$ to $-\infty$ &  
$\pm\infty$ & 
$\infty$ to $1$ \\

$H_{\mbox{\scriptsize min}}$
($\frac{\delta^2}{2P_0}$) &
$-1$ to $-1/2$ &
$-1/2$ &
$-1/2$ to $0$ &
$0$ to $\infty$ &
$\pm\infty$ &
$-\infty$ to $-1$ \\

sign of $F$  & $+$  & $+$  &   $+$  &    $-$  &   $\infty$   &   $+$ \\

sign of $\Ft$& $+$  & $+$  &   $+$  &    $+$  &   $\infty$   &   $-$ \\

type of system & 
Greg. &
eq. Greg. &
inv. Greg. &
Cass. &
eq. Cass.  &
inv. Cass. \\
\tableline
\end{tabular}
\caption{
Characteristics of 2-mirror afocal systems, generated 
from the amplitude mapping function $A(\rt) \equiv \alpha$.
Here, {\em inv.} means inverted, {\em eq.} means equal and 
refers to relative mirror sizes, {\em Greg.} means Gregorian, and {\em Cass.}
means Cassegrain.  Note that the ``equal Cass.'' is a plane-mirror periscope.
}
\label{tbl1}
\end{table}
\vfill
 
\end{document}